\documentclass[12pt,thmsa]{article}
\usepackage{sw20lart}

\input tcilatex

\begin{document}

\author{I. Radinschi\thanks{%
radinschi@yahoo.com, iradinsc@phys.tuiasi.ro} \and Physics Department, ``Gh.
Asachi'' Technical University, \and Iasi, 700050, Romania}
\title{KKW Analysis for the Dyadosphere of a Charged Black Hole}
\date{November 28, 2005}
\maketitle

\begin{abstract}
The Keski-Vakkuri, Kraus and Wilczek (KKW) analysis is used to compute the
temperature and entropy in the dyadosphere of a charged black hole solution.
For our purpose we choose the dyadosphere region of the Reissner-Nordstr\"{o}%
m black hole solution.

Our results show that the expressions of the temperature and entropy in the
dyadosphere of this charged black hole are not the Hawking temperature and
the Bekenstein-Hawking entropy, respectively.


PACS: 04. 20 Dw, 04. 70. Bw,
\end{abstract}

\section{INTRODUCTION}

One of the most interesting problem of relativity is the evaluation of the
temperature and entropy of black holes and this issue still attracts
considerable attention in the literature. The important method of
Keski-Vakkuri, Kraus and Wilczek (KKW) [1] has been used to compute the
temperature and entropy of Schwarzschild-type black hole solution [2] and,
after this, has been applied to other black hole space-times [3]-[5]. It is
important to point out that in the (KKW) analysis, the total
Arnowitt-Desser-Misner mass [6] is fixed but the mass of the Schwarzschild
black hole decreases due to the emitted radiation. For solving the problem
of singularities and to avoid them at the horizon the Painlev\'{e} [7]
coordinate transformation is used and this enable us the study of the
across-horizon physics such as the black hole radiation. The black hole
temperature depends not only on the characteristics of the black hole but,
also, on the energy of the emitted shell of energy. Furthermore, the black
hole entropy is not given by Bekenstein and Hawking formula for the specific
black hole.

Also, the (KKW)\ generalized analysis can be applied successfully for
evaluating the temperature and entropy of a black hole solution which is not
of Schwarzschild-type. The generalized (KKW) analysis case was studied by
Vagenas [8] and he gave the formulas for the temperature and entropy of a
black hole solution described by a metric which satisfies the condition $%
A(r)\cdot B^{-1}(r)\neq 1$. About the Hawking radiation, we are allowed to
conclude that is viewed as a tunneling process which emanates from the
non-Schwarzschild-type black hole solution.

Vagenas [8] used a more general coordinate transformation in order to apply
the (KKW)\ analysis to non-Schwarzschild-type black holes. The two
conditions: 1) the regularity at the horizon which ensures that we can study
the across-horizon physics and 2) the stationarity of the non-static metric
which implies that the time direction is a Killing vector, were fulfilled in
order to generalize the (KKW) analysis. This generalized analysis furnishes
us the exact expressions of the temperature and entropy of the
non-Schwarzschild-type black holes which are not the Hawking temperature $%
T_{H}$ and the Bekenstein-Hawking entropy $S_{BH}$. Other interesting
results on black hole thermodynamics were obtained [9]. Furthermore, we use
the (KKW)\ generalized analysis in order to compute the temperature and
entropy of a magnetic stringy black hole solution [10].

In this paper we use the Keski-Vakkuri, Kraus and Wilczek (KKW) analysis to
compute the temperature and entropy in the dyadosphere of a charged black
hole solution. We choose the dyadosphere region [11] of the Reissner-Nordstr%
\"{o}m black hole solution .

\section{KKW ANALYSIS FOR THE DYADOSPHERE\ OF\ A\ CHARGED BLACK\ HOLE}

Ruffini and collaborators [12]-[13] demonstrated that the event horizon of a
charged black hole is surrounded by a special region called the dyadosphere
where the electromagnetic field exceeds the Euler-Heisenberg critical value
for electron-positron pair production. They studied certain properties of
the dyadosphere corresponding to the Reissner-Nordstr\"{o}m space-time
[12]-[13]. In addition, the new concept of dyadosphere was introduced by
Ruffini [12] to explain gamma ray bursts.

De Lorenci, Figueiredo, Fliche and Novello [11] computed the correction for
the Reissner-Nordstr\"{o}m metric from the first contribution of the
Euler-Heisenberg Lagrangian and got the metric

\begin{eqnarray}
ds^{2} &=&-(1-\frac{2\,M}{r}\,+\frac{Q^{2}}{r^{2}}-\frac{\sigma \,Q^{4}}{%
5\,r^{6}})\,dt^{2}+  \TCItag{1} \\
&&+(1-\frac{2\,M}{r}\,+\frac{Q^{2}}{r^{2}}-\frac{\sigma \,Q^{4}}{5\,r^{6}}%
)\,^{-1}\,dr^{2}+r^{2}\,\,d\theta ^{2}+r^{2}\,\sin ^{2}\theta \,d\varphi
^{2}.  \nonumber \\
&&  \nonumber
\end{eqnarray}

This metric is of the type

\begin{equation}
ds^{2}=-A(r)\,dt^{2}+A(r)\,^{-1}\,dr^{2}+r^{2}\,\,d\theta ^{2}+r^{2}\,\sin
^{2}\theta \,d\varphi ^{2},  \tag{2}
\end{equation}

with $A(r)=1-\frac{2\,M}r\,+\frac{Q^2}{r^2}-\frac{\sigma \,Q^4}{5\,r^6}$.

For $\sigma =0$ we obtain the Reissner-Nordstr\"{o}m solution. De Lorenci 
\textit{et. al}. [11] showed that the correction term $\frac{\sigma \,Q^4}{%
5\,r^6}$ is of the same order of magnitude as the Reissner-Nordstr\"{o}m
charge term $\frac{Q^2}{r^2}$. The metric describes a black hole with an
event horizon at $r_{+}$.

The (KKW) methodology [1] in the case of a black hole background which
belongs to the family of geometries of Schwarzschild-type requires that the
metric should be regular at the event horizon. Also, the total
Arnowitt-Desser-Misner mass $M_{ADM}$ have to be well-defined so we get $%
A(r)\rightarrow 1,$ as $r\rightarrow \infty $. We have the Painlev\'{e} [7]
coordinate transformation given by

\begin{equation}
\sqrt{A(r)}dt=\sqrt{A(r)}d\tau -\sqrt{\frac{1-A(r)}{A(r)}}dr,  \tag{3}
\end{equation}

where $\tau $ is the new time coordinate. The metric given by (1) becomes

\begin{equation}
ds^{2}=-A(r)\,d\tau ^{2}+2\sqrt{1-A(r)}\,d\tau \,dr+dr^{2}+r^{2}\,d\theta
^{2}+r^{2}\,\sin ^{2}\theta \,d\varphi ^{2}.  \tag{4}
\end{equation}

The radial null geodesics are

\begin{equation}
\QATOP{.}{r}=_{-}^{+}1-\sqrt{1-A(r)}.  \tag{5}
\end{equation}

In the equation above, the upper (lower) sign corresponds to the outgoing
(ingoing) geodesics under the assumption that $\tau $ increases towards
future.

The total Arnowitt-Desser-Misner mass $M_{ADM}$ is fixed and the mass $M$ of
the black hole fluctuates because a shell of energy $\omega $ which
constitutes of massless particle considering only the s-wave part of
emission, is radiated by the black hole. Now, we are in the situation when
the massless particles travel on the outgoing geodesics which are due to the
varying mass $M$ of the black hole. The metric becomes

\begin{equation}
\begin{tabular}{c}
$ds^{2}=-A(r,M-\omega )\,d\tau ^{2}+2\sqrt{1-A(r,M-\omega )}\,d\tau \,dr+$
\\ 
\\ 
$+dr^{2}+r^{2}\,d\theta ^{2}+r^{2}\,\sin ^{2}\theta \,d\varphi ^{2}.$%
\end{tabular}
\tag{6}
\end{equation}

We get for the outgoing radial null geodesics a new formula

\begin{equation}
\QATOP{.}{r}=1-\sqrt{1-A(r,M-\omega )}.  \tag{7}
\end{equation}

We make the approximation

\begin{equation}
\sqrt{1-A^{^{\prime }}}\approx 1-\frac{1}{2}A^{^{\prime }},  \tag{8}
\end{equation}

where $A^{^{\prime }}=A(r,M-\omega ^{^{\prime }})\,$and and thus the
imaginary part of the action (see equation (19) in [8])

\begin{equation}
\func{Im}I=\func{Im}\int_{r_{+}(M-\omega )}^{r_{+}(M}\int_{0}^{+\omega }%
\frac{d\omega ^{^{\prime }}}{1-\sqrt{1-A^{\prime }}}dr.  \tag{9}
\end{equation}

We obtain for the metric given by (1)

\begin{equation}
\func{Im}I=\frac{\pi }{2}[r_{+}^{2}(M)-r_{+}^{2}(M-\omega )].  \tag{10}
\end{equation}

Now, we evaluate the temperature of the black hole (see equation (20) in
[8]) and we obtain

\begin{equation}
T_{bh}(M,\omega )=\frac{\omega }{\pi }[r_{+}^{2}(M)-r_{+}^{2}(M-\omega
)]^{-1}.  \tag{11}
\end{equation}

The expression of the entropy is given by

\begin{equation}
S_{bh}=S_{BH}-\pi \lbrack r_{+}^{2}(M)-r_{+}^{2}(M-\omega )].  \tag{12}
\end{equation}

The Hawking temperature $T_H$ in the dyadosphere of the charged black hole
is defined as

\begin{equation}
T_{H}=\frac{r_{+}-r_{-}}{A_{h}}=\frac{r_{+}-r_{-}}{4\,\pi r_{+}^{2}}. 
\tag{13}
\end{equation}

The entropy of the black hole is different from the Bekenstein-Hawking
entropy formula $S_{BH}$ that is given by

\begin{equation}
S_{BH}=\frac{A_{h}}{4}=\pi \,r_{+}^{2}.  \tag{14}
\end{equation}

\section{DISCUSSION}

One of the most attractive methods used to evaluate the temperature and
entropy of black holes is the (KKW) analysis. In some recent inestigations
about this important issue, the temperature and entropy of black holes, the
importance of the (KKW) analysis is emphasized [14]. An interesting study
about the Hawking radiation and the temperature and entropy of black holes
was made by M. Angheben et. al. [15], and A. J. M. Medved and E. C. Vagenas
[16]. In these works, the authors also pointed out some interesting results
obtained with the (KKW) analysis.

We used the (KKW) analysis introduced in [1] in order to evaluate the
temperature and entropy in the dyadosphere of the Reissner-Nordstr\"{o}m
black hole solution. We conclude that the temperature and entropy in the
dyadosphere region of the Reissner-Nordstr\"{o}m black hole solution are
different from the Hawking temperature $T_{H}$ and the Bekenstein-Hawking
entropy $S_{BH}$, respectively. The temperature and entropy in the
dyadosphere depend on the $r_{+}$ and $r_{-}$ and on $r_{+}$ parameters,
respectively. Since we don't know the explicit expression of $r_{+}$ and $%
r_{-}$, we are allowed only to assume that these expressions depend on the
emitted particle's energy. Our results sustain the importance of the KKW
analysis [1].

\end{document}